\documentclass[a4paper]{article}

\usepackage{INTERSPEECH2020}

\usepackage{anyfontsize}
\usepackage{t1enc}

\usepackage{graphicx}
\usepackage{amssymb,amsmath,bm}
\usepackage{textcomp}
\usepackage{color}
\usepackage{multirow}
\usepackage{hyperref}
\usepackage{url}
\usepackage[activate]{microtype}
\usepackage{subcaption}
\usepackage{soul}
\usepackage{cleveref}

\def\vec#1{\ensuremath{\bm{{#1}}}}

\newcommand{\audeep}{\mbox{\textsc{auDeep }}}
\newcommand{\compare}{\mbox{\textsc{ComParE }}}

\PassOptionsToPackage{dvipsnames,table}{xcolor}
\usepackage{pgfplotstable}
\pgfplotsset{compat=1.14}

\ninept

\newcommand{\eg}{e.\,g., }
\newcommand{\ie}{i.\,e., }

\newcommand{\etal}{et\,al.}

\title{A Novel Fusion of Attention and Sequence to Sequence Autoencoders to Predict Sleepiness From Speech}

\name{Shahin Amiriparian$^{1}$, Pawel Winokurow$^{1}$, Vincent Karas$^{1,3}$, Sandra Ottl$^{1}$, Maurice Gerczuk$^{1}$, \\Bj\"{o}rn W.\ Schuller$^{1,2}$}
\address{\fontsize{11}{11}\selectfont 
    $^1$EIHW -- Chair of Embedded Intelligence for Health Care and Wellbeing, University of Augsburg, Germany \\
    $^2$GLAM -- Group on Language, Audio, \& Music, Imperial College London, UK \\
    $^3$BMW AG}
\email{amiriparian@IEEE.org}

\begin{document}

\maketitle
\begin{abstract}
Motivated by the attention mechanism of the human visual system and recent developments in the field of machine translation, we introduce our attention-based and recurrent sequence to sequence autoencoders for fully unsupervised representation learning from audio files. In particular, we test the efficacy of our novel approach on the task of speech-based sleepiness recognition. We evaluate the learnt representations from both autoencoders, and conduct an early fusion to ascertain possible complementarity between them. In our frameworks, we first extract Mel-spectrograms from raw audio. Second, we train recurrent autoencoders on these spectrograms which are considered as time-dependent frequency vectors. Afterwards, we extract the activations of specific fully connected layers of the autoencoders which represent the learnt features of spectrograms for the corresponding audio instances. Finally, we train support vector regressors on these representations to obtain the predictions. On the development partition of the data, we achieve Spearman's correlation coefficients of $.324$, $.283$, and $.320$ with the targets on the Karolinska Sleepiness Scale by utilising attention and non-attention autoencoders, and the fusion of both autoencoders' representations, respectively. In the same order, we achieve $.311$, $.359$, and $.367$ Spearman's correlation coefficients on the test data, indicating the suitability of our proposed fusion strategy.

\end{abstract}

\section{Introduction}
\label{sec:Introduction}
Sleepiness is a state identified by reduced alertness that varies according to a circadian rhythm, \ie with the time of the day~\cite{Dement.1982,Killgore.2010}. Its detection is important for safety applications, as it has been shown, for example, that sleepiness impacts driving performance, even more so than fatigue~\cite{Philip.2005}. 
\cite{lyznicki1998sleepiness}
Most systems that aim to detect a sleepy driver rely on signals derived from interaction with the vehicle, such as abnormal steering behaviour, failures in lane keeping or irregular use of the pedals~\cite{Mashko.2015}. Research dealing with automatic sleepiness recognition has investigated methods to derive the state based on different bio-signals. Performing visual analysis of a subject's face, \eg measuring  blinking, can serve in assessing sleepiness but may be negatively affected by changing environmental parameters, such as illumination~\cite{F.Friedrichs.2010}. While electroencephalography (EEG)  has also been shown to be a robust approach to the problem~\cite{Balandong.2018}, it is far more intrusive and can only be achieved with special equipment and professional setup. In contrast to this, performing acoustic analysis of speech is non-obtrusive and does not need as much sensor application and calibration effort~\cite{Krajewski09-ASD}. 

In order to define a suitable target for the automatic analysis of sleepiness, the state can be described by the Karolinska Sleepiness Scale (KSS) in terms of ratings ranging from 1 to 9, sometimes additionally extended by an extra category 10 for extreme sleepiness. A binary classification of sleepy speech has been previously performed as part of the INTERSPEECH 2011 Speaker State Challenge~\cite{Schuller11-TI2}. The subchallenge of INTERSPEECH 2019's \compare challenge deals with the detection of continuous sleepiness, and the sleepiness of a speaker is assessed as a regression problem~\cite{Schuller19-TI2}. In the challenge baseline, the problem was approached in a number of ways, including a traditional acoustic feature extraction pipeline, Bag-of-Audio-Words, and a deep recurrent autoencoder framework. Here, the unsupervised sequence to sequence model \audeep achieved the strongest results. 

Deep learning models, such as \audeep, that process raw or low level inputs yield state-of-the-art results for a wide range of machine learning problems~\cite{young2018recent, amodei2016deep, he2016deep}. Many of these approaches consider inputs as a whole, treating every part with the same importance. Often, however, some pieces of the input contain more information pertinent to solving the task at hand than others. A popular approach that takes this notion into consideration can be found with attention mechanisms, such as the one introduced by Bahdanau~\etal~\cite{bahdanau} for machine translation. Compared to regular sequence to sequence autoencoders where all information is compressed into the last hidden state of the encoder, the dynamic context vector in the attention model retains information about all hidden states of the encoder and their alignment to the current decoding step. Since their introduction, attention mechanisms have also been adapted to speech recognition~\cite{chorowski2015attention, bahdanau2016end}, visual image captioning~\cite{xu2015show} or question answering~\cite{anderson2018bottom}, and speech emotion classification~\cite{huang2017deep}. 

Motivated both by the effectiveness of recurrent autoencoders for acoustic analysis of sleepiness from speech as well as by the improvements to sequence to sequence models achieved with attention mechanisms, we evaluate the impacts of combining the two approaches for the detection of continuous sleepiness on the respective 2019 INTERSPEECH \compare sub-challenge.

\begin{figure*}[ht!]
    \centering
    \includegraphics[width=.85\linewidth]{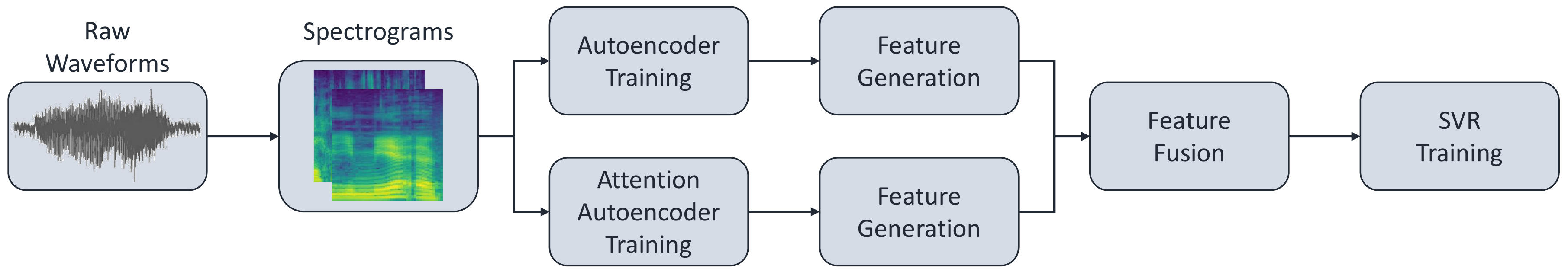}
    \caption{An overview of our framework for unsupervised representation learning with sequence to sequence recurrent autoencoders. Except for the SVR training, the approach is fully unsupervised. A detailed description of the procedure is given in~\Cref{sec:approach}.}
    \label{fig:sys-overview-s2s}
\end{figure*}
\section{Dataset}\label{sec:dataset}
We use a subset of the SLEEP Corpus that was employed in the 2019 edition of the INTERSPEECH Computational Paralinguistics Challenge (\compare)~\cite{Schuller19-TI2}. The corpus contains speech recordings of 915 individuals (364 females, 551 males) at varying levels of sleepiness. The audio files have a sample rate of 16\,kHz with a 16\,bit quantisation. Each file is annotated with a KSS-score, ranging from 1 to 9, with 9 denoting extreme sleepiness. The labels were derived by averaging self-report with the scores of two external observers. The dataset is split into three partitions with 5\,564, 5\,328, and 5\,570 samples.

\section{Approach}\label{sec:approach}
A high-level overview of our proposed approach is depicted in~\Cref{fig:sys-overview-s2s}. First, Mel-spectrograms are extracted from audio signals. Recurrent autoencoders (AEs), both with and without attention mechanism, are then trained on the Mel-spectrograms. Afterwards, the learnt representations are obtained from the AEs. We then fuse the representations and classify them.

\subsection{Spectrogram Extraction}\label{ssec:spectrogram-extraction}
First, the Mel-spectrograms of audio recordings are extracted using periodic Hamming windows with width $w$ and overlap $0.5w$. From these, a given number of log-scaled Mel-frequency
bands are then computed. Finally, we normalise the Mel-spectra to have values in [-1; 1], since the outputs of the recurrent sequence to sequence autoencoders are constrained to this interval.

\subsection{Autoencoder Architecture without Attention}\label{ssec:ae-wo-att}
For this architecture, we utilise \textsc{auDeep}\footnote{\url{https://github.com/auDeep/auDeep}}~\cite{Amiriparian17-STS,Freitag18-AUL}, our recurrent sequence to sequence autoencoder. For the representation learning with our framework, we can adjust a range of autoencoder parameters, including the direction (\eg uni- or bidirectional) of the encoder and decoder RNNs, types of RNN cells, \eg gated recurrent units (GRUs), or long short-term memory (LSTM) cells, and the number of hidden layers and units. To use LSTM-RNNs in the decoder, the LSTM cell is modified to work with a context vector similar to the GRUs in the encoder-decoder model proposed by Cho~\etal~\cite{seq2seq_cho}. The weight matrices $C_i, C_f, C_o$, and $C_z$ are added to the input $z_t$, input gate $i_t$, forget gate $f_t$, and output gate $o_t$ to enable an LSTM 
cell 
to work with the context vector:
\begin{equation}
\begin{aligned}
 z_t &= \tanh \left(W_zx_t+R_z y_{t-1} + C_z c + b_z \right)  \\
i_t &= \sigma \left(W_ix_t + R_iy_{t-1} + C_i c + p_i \odot c_{t-1} + b_i\right)\\
f_t &= \sigma \left(W_fx_t + R_fy_{t-1} + C_f c + p_f \odot c_{t-1} + b_f\right)\\
o_t &= \sigma \left(W_ox_t + R_oy_{t-1} + C_o c + p_o \odot c_t + b_o\right).
\end{aligned}
\end{equation}

For each input sequence, the initial hidden state vector of the encoder is zero-padded. The last concatenated hidden state vector of the encoder $h_T^{e} = \left[\overrightarrow{h}_{T} \overleftarrow{h}_{T}\right]^T$ is then passed through a fully connected layer with $\tanh$ activation which has the same number of units as the decoder RNN. The output of this layer represents the context vector and is used as the first hidden state vector of the decoder $h^{d}_0$. During the feature extraction, the context vector also represents the feature vector.
The outputs of the decoder are passed through a fully connected projection layer with $\tanh$ activation at each time step in order to map the decoder output dimensionality to the target dimensionality. The weights of this output projection are shared across time steps.
For the network training the teacher forcing algorithm~\cite{lamb2016professor} is applied. Following this method, instead of feeding the decoder with the predicted output at time step $t-1$ ($\hat{y}_{t-1}$), the expected decoder output at time step $t-1$ ($y_{t-1}$) is fed as an input to the decoder.
This means that the decoder input is the same original spectrogram, only shifted by one step in time. Instead of the first step, the zero vector is inserted and the frequency vector at the last step is removed. 
During the training, the autoencoder learns to reconstruct the reversed input spectrogram~\cite{Amiriparian17-STS, seq2seq_sutskever}. Mean squared error (MSE) is used as the loss function to compare the reversed source spectrogram with the concatenated spectrogram obtained from the projection layer.

\begin{figure*}[tph!]
\centering
\includegraphics[width=.90\linewidth]{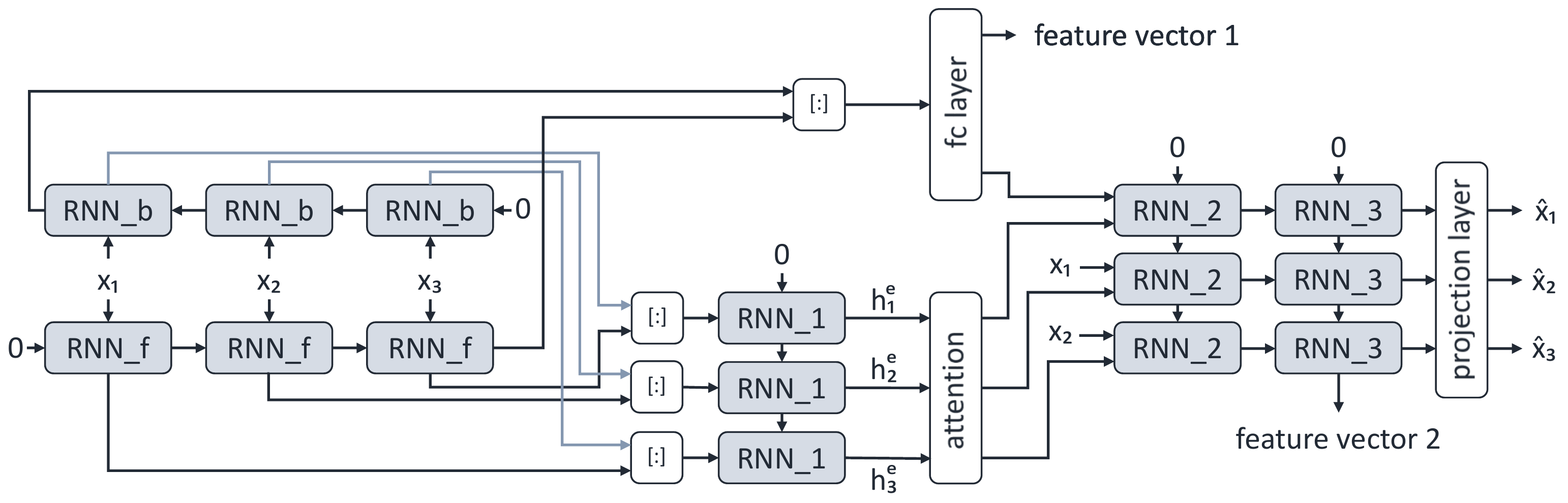}
\caption{Schematic structure of our attention-based autoencoder with stacked encoder and decoder RNNs. Feature vectors 1 and 2 are extracted from the activations of the fully connected layer of the encoder RNN and the last hidden state of the decoder RNN, respectively. A detailed account of the proposed architecture is provided in~\Cref{ssec:ae-w-att}.} \label{fig:att-model}
\end{figure*}

\subsection{Autoencoder Architecture With Attention}\label{ssec:ae-w-att}

In the second model, we add an attention mechanism to the autoencoder architecture. Here, encoder and decoder have almost the same structure as the baseline autoencoder. The problem of the sequence to sequence model is that the encoder must map all essential information of the input sequence to a fixed-length vector. This may not be enough to represent a long input sequence. To circumvent this, Bahdanau \etal~\cite{bahdanau} have introduced an attention mechanism that includes dynamic computation of the context vector. At each time step of the decoder, the attention mechanism enables to choose the hidden state vectors of the encoder that contain the most significant information to generate the context vector $c_t$ which is used to generate the output $y_t$. This computation is based on all hidden state vectors of the encoder and the last hidden state vectors of the decoder. The context vector $c_i$ is the linear combination of the hidden state vectors of the encoder $h^{e}_j$:
\begin{gather}
	c_i = \sum_{j=1}^{T_x} \alpha_{ij} h_j^{e}. 
\end{gather}
Here, the weights $\alpha_{ij}$ are numbers between 0 and 1 and define which hidden states $h^{e}_j$ have the biggest influence on $y_i$. $\alpha_{ij}$ is calculated with the softmax-normalised inner activation of the alignment model $e_{ij}$:
\begin{gather}
	\alpha_{ij}=\frac{\exp({e_{ij}})}{\sum_{k=1}^{T_x} \exp({e_{ik}})}. 
\end{gather}
The alignment model $e_{ij} = a(h^e_{i-1}, h^d_j)$ is a small feedforward neural network trained together with the sequence to sequence model using backpropagation and is defined as
\begin{gather}
    v^\top_{a_i}\tanh \left( W_a h^{d}_{i-1} + U_a h^e_j \right),
\end{gather}
where $v_a\in\mathbb{R}^l$ is a weight vector, and $W_a\in\mathbb{R}^{l\times n}$ and $U_a\in\mathbb{R}^{l\times 2n}$ are weight matrices, and $l$ is the number of units in the alignment model.

The difference to the autoencoder described in \Cref{ssec:ae-wo-att} is that the context vector is calculated dynamically using the alignment model and is not solely used as a feature vector. If we compute the feature vector from the last hidden state vector of the encoder, the attention mechanism will have less influence on the feature extraction. The feature vector, however, should contain information of all context vectors and, therefore, is extracted after their creation. For this reason, an additional set of feature vectors is extracted from the hidden state vector of the last decoder layer at the last time-step. The attention mechanism in the sequence to sequence model is visualised in~\Cref{fig:att-model}.

\section{Experimental Settings}\label{sec:experimental-settings}
For all experiments, we train autoencoder models on the sleepiness spectrograms and then extract features for the three partitions. These feature vectors are used to train a linear support vector regressor (SVR) for which we optimise the complexity parameter on a logarithmic scale of $10^{-5}$ to 1. The complexity parameter is chosen based on the Spearman's correlation coefficient ($\rho$) achieved on the development partition.

Our proposed autoencoder approaches contain a large amount of adjustable hyperparameters (cf.~\Cref{ssec:ae-w-att,ssec:ae-wo-att}), which prohibits an exhaustive exploration of the parameter space. For this reason, we choose suitable values for the hyperparameters in multiple stages, using the results of our initial experiments to bootstrap the process. In preliminary experiments, we test different configurations for the spectrograms that are used as input for the autoencoders. We arrived at using Mel-spectrograms with 160 and \{128, 256\} Mel-bands extracted from the audio samples for our autoencoders with and without attention, respectively. 
For the fast Fourier transform (FFT), we apply Hamming windows of 40\,ms width and 20\,ms overlap. We further experimented with the architecture of our recurrent autoencoder models. Here, we tested one and two layer variants for both encoder and decoder using either GRU or LSTM cells with 128, 256, or 512 hidden units. Moreover, bidirectional and unidirectional encoders were compared. For the attention autoencoder, models with two-layer bidirectional encoders and two-layer unidirectional decoders worked best with 512 hidden units. For the results presented herein, we therefore settled on this architecture with the choice of either GRU or LSTM layers (cf.~\Cref{tab:attention-results}). Furthermore, we introduce an additional RNN layer in the decoder that serves to produce hidden states for the attention mechanism. The architecture of those models is visualised in~\Cref{fig:att-model}. We evaluate features extracted from both the last hidden state of the encoder and decoder. All attention models are trained using a batch size of 256 with the Adam optimiser~\cite{adam} and the learning rate set to $10^{-4}$ for a maximum of 40 epochs. The model checkpoints at 20, 25, 30, 35, and 40 epochs further serve as feature extractors.
For the \audeep experiments (cf.~\Cref{ssec:ae-wo-att}), we found the best configuration with 2 hidden layers each with 256 hidden units. We then optimise the direction of the encoder and decoder and adjust the RNN cell type. Additionally, we filter some of the background noise in the recordings by clipping amplitudes below \{-40, -50, -60, -70\}\,dB thresholds, and fuse them together resulting in five different feature vectors for each data partition. In~\cite{Amiriparian18-DUR,Amiriparian18-AFO}, we have shown the effectiveness of our amplitude clipping approach for various audio processing tasks.

\section{Results and Discussion}
\label{sec:results}
All results obtained with our autoencoders and their early fusion are shown in~\Cref{tab:attention-results,tab:audeep-results,tab:fusion-results}. In the attention model (cf.~\Cref{tab:attention-results}), features from the fully connected layer of the encoder RNN (fc$_{enc}$) generalise better when GRUs are applied ($\rho_{devel} = .250$, $\rho_{test} = .314$), and the features from the last hidden state of the decoder RNN (state$_{dec}$) perform better with LSTM cells ($\rho_{devel} = .324$, $\rho_{test} = .311$). 
From both autoencoder approaches, the recurrent model without attention shows the best performance on the test partition ($\rho_{test} = .359$), whilst the attention model achieves the highest results on the development partition ($\rho_{devel} = .324$). The result 
implies 
possible overfitting of the attention model on the development data. This issue is not strongly present in our model without attention, and we hypothesise that this is mainly because of the filtering of some of the background noise found in the audio data by clipping amplitudes below a certain threshold. In~\Cref{tab:audeep-results}, we provide the highest achieved results with various thresholds and hyperparameter combinations. Furthermore, we fuse the best performing attention feature set on the development set ($\rho_{devel} = .324$, $\rho_{test} = .311$) with all non-attention (\textsc{auDeep}) features to analyse the complementarity of the learnt representations. The results in~\Cref{tab:fusion-results} demonstrate an improvement of all results after early fusion. The highest improvement on the test partition after fusion is achieved when the best attention feature set is combined with the fourth \audeep feature with GRUs, 
and the 
unidirectional encoder and bidirectional encoder trained on Mel-spectrograms with 256 Mel-band (FFT window width of 80\,ms and overlap of 40\,ms) and -60\,dB amplitude clipping (cf.~\Cref{tab:fusion-results}). It is worth mentioning that the dimensionality of the attention features are either $1/2$ or $1/8$ of the \audeep features, leading to a faster classifier training. Moreover, the training process with attention autoencoders can be performed faster, as the encoder RNN is relieved from encoding all information in the whole input sequence of the Mel-spectrograms into a fixed-length vector~\cite{Amiriparian19-DRL}.
We further compare our best performing approaches with 
the 
best challenge baselines~\cite{Schuller19-TI2}, the winner of the challenge who combined Fisher vectors with baseline features~\cite{Gosztolya.2019}, and the runner-up who utilised a fusion of convolutional neural networks (CNNs) and RNNs~\cite{yeh2019using} (cf.~\Cref{tab:compare_results}).

\begin{table}[tp]
    \centering
    \caption{Performance comparison of the features obtained from the fully connected layer of the encoder (fc$_{enc}$) and the last hidden state of the decoder (state$_{dec}$) in our attention autoencoder. id: feature identifier, Dim.: feature dimensionality.}
    \resizebox{\columnwidth}{!}{
    \begin{tabular}{rrlrrrrr}
        \toprule
        \multicolumn{4}{c}{Parameters} & \multicolumn{2}{c}{fc$_{enc}$} &  \multicolumn{2}{c}{state$_{dec}$} \\
        \midrule
        id & Epoch & Cell & Dim. & $\rho_{devel} $ & $\rho_{test}$ & $\rho_{devel} $ & $\rho_{test}$ \\
        \midrule
        1 & 20 & GRU    & 512 & .262 & .308 & .276 & .294 \\
        2 & 25 & GRU    & 512 & .258 & .308 & .278 & .298  \\
        3 & 30 & GRU    & 512 & .250 & .314 & .267 & .292  \\
        4 & 35 & GRU    & 512 & .260  & .312 & .266 & .298  \\
        5 & 40 & GRU    & 512 & .253  & .307 & .265 & .288  \\
        \midrule
        6 & 20 & LSTM   & 512 & .293 & .294 & .318 & .303 \\
        7 & 25 & LSTM   & 512 & .303 & .276 & \textbf{.324} & \textbf{.311} \\
        8 & 30 & LSTM   & 512 & .298 & .263 & .322 & .305 \\
        9 & 35 & LSTM   & 512 & .303 & .285 & .289 & .304 \\
        10 & 40 & LSTM   & 512 & .307 & .302 & .288 & .299 \\
        \bottomrule
    \end{tabular}
    }
    \label{tab:attention-results}
\end{table}

\begin{table}[tp]
    \centering
    \caption{Results obtained from our autoencoder without attention. id: feature identifier, w: width of the Hamming window, Mel: number of Mel-bands, Clip: clipped amplitudes below a certain threshold to filter some noise from audio, Dim.: feature dimensionality, Dir.: direction of the encoder-decoder.}

    \resizebox{\columnwidth}{!}{
    \begin{tabular}{rrrrrrrrr}
        \toprule
        \multicolumn{8}{c}{Autoencoders without attention} \\
        \midrule
        id & w & Mel & Clip & Cell & Dim. & Dir. & {$\rho_{devel}$} & {$\rho_{test}$} \\
        \midrule
        1 & 0.08  & 256 & fused & GRU & 4\,096 & uni-bi & .286 & .338 \\
        2 & 0.08  & 256 & -70\,dB   & GRU & 1\,024 &  bi-uni & \textbf{.283} & \textbf{.359}\\
        3 & 0.06  & 256 & -70\,dB   & GRU & 1\,024 &  bi-uni & .281 & .357 \\
        4 & 0.08  & 256 & -60\,dB   & GRU & 1\,024 &  uni-bi & .278 & .331 \\
        5 & 0.04  & 256 & -70\,dB   & GRU & 1\,024 &  uni-bi & .278 & .340 \\
        6 & 0.06  & 256 & fused & GRU & 4\,096 &  bi-uni & .277 & .346 \\
        7 & 0.06  & 256 & -70\,dB   & GRU & 1\,024 &  uni-bi & .277 & .348 \\
        8 & 0.06  & 128 & -70\,dB   & LSTM & 1\,024 &  uni-bi & .277 & .317 \\
        9 & 0.08  & 128 & -60\,dB   & GRU & 1\,024 &  bi-uni & .275 & .324 \\
        10 & 0.04  & 256 & -60\,dB   & GRU & 1\,024 &  uni-bi & .275 & .336 \\
         \bottomrule
    \end{tabular}
    }
    \label{tab:audeep-results}
\end{table}{}

\section{Conclusions and Future Work}
\label{sec:conclusions}
In~\Cref{ssec:ae-wo-att}, we introduced a novel attention mechanism for recurrent sequence to sequence autoencoders to process audio signals in an unsupervised manner\footnote{Our audio-based attention framework (\textsc{auttention}), and all codes to reproduce the attention results are provided here: \mbox{\url{https://github.com/auttention/SleepyAttention}}}. We have demonstrated the suitability of two fully unsupervised representation learning techniques for continuous sleepiness recognition, and their superior performance to expert-designed, hand-crafted features (cf.~\Cref{tab:compare_results}). Furthermore, we conducted a feature fusion strategy and demonstrated the complementarity of the learnt representations from both autoencoder architectures (cf.~\Cref{sec:results}). In future work, we will be evaluating our systems over a wide range of audio recognition tasks. We also want to utilise dimensionality reduction techniques~\cite{Amiriparian17-FSI} to cope with the high-dimensionality of our autoencoder features. Finally, we want to combine our methods with CNN-based representation learning systems, such as deep convolutional generative adversarial networks~\cite{radford2015unsupervised}.

\begin{table}[tp]
    \centering
    \caption{Results of our early fusion experiments with the best attention result (id$_{att}$ = 7) and all results provided in~\Cref{tab:audeep-results}. id$_{att}$ and id$_{audeep}$: identifiers for the attention feature and \audeep features which are fused. $C_{SVR}$: Complexity of the SVR which is optimised on the development partition after fusion.}
    \begin{tabular}{rr|rrr}
        \toprule
        \multicolumn{5}{c}{Early fusion} \\
        \midrule
        id$_{att}$ &
        id$_{audeep}$ &
        $C_{SVR}$ &
        {$\rho_{devel}$} &
        {$\rho_{test}$} \\
        \midrule
        7 & 1 & $10^{-3}$ & .315 & .359 \\
        7 & 2 & $10^{-2}$ & .336 & .360 \\
        7 & 3 & $10^{-2}$ & .334 & .365 \\
        7 & 4 & $10^{-1}$ & \textbf{.320} & \textbf{.367} \\
        7 & 5 & $10^{-2}$ & .333 & .349 \\
        7 & 6 & $10^{-3}$ & .319 & .363 \\
        7 & 7 & $10^{-2}$ & .326 & .361 \\
        7 & 8 & $10^{-2}$ & .333 & .341 \\
        7 & 9 & $10^{-2}$ & .340 & .351 \\
        7 & 10 & $10^{-2}$ & .339 & .357 \\
         \bottomrule
    \end{tabular}
    \label{tab:fusion-results}
\end{table}{}

\begin{table}[tp]
    \centering
    \caption{Comparison of our best performing models with best performing challenge baselines and the challenge winner. $Dim.$: feature dimensionality of each system.}
    \resizebox{\columnwidth}{!}{
    \begin{tabular}{lrrr}
        \toprule
        System & Dim. & $\rho_{dev}$ & $\rho_{test}$ \\
        \midrule
        \multicolumn{4}{l}{\textbf{Challenge Winner}~\cite{Gosztolya.2019}} \\
        \compare + BoAW + Fisher vectors & -- & -- & \textbf{.383} \\
        \multicolumn{4}{l}{\textbf{Runner-up}~\cite{yeh2019using}} \\
        CNNs and BLSTMs with attention & -- & .373 & .369 \\ 
        
        \midrule
        \multicolumn{4}{l}{\textbf{Best Challenge Baselines}~\cite{Schuller19-TI2}} \\
        \compare                            & 6\,373    & .251  & .314 \\
        Bag-of-Audio-Words                  &  500      & .250  & .304 \\
        Autoencoders                        & 1\,024    & .243  & .325 \\
        Late fusion of best                 & --         & --     & .343 \\
        \midrule
        \multicolumn{4}{l}{\textbf{Best of Our Proposed Approaches}} \\
        With attention                      & 512       & .324  & .311 \\
        Without attention                   & 4\,096    &  .286 & .338 \\
        Early fusion                        & 4\,608    & \textbf{.320}  & \textbf{.367} \\
        \bottomrule
    \end{tabular}
    }
    \label{tab:compare_results}
\end{table}{}
\section{Acknowledgements}
This research was partially supported by BMW AG and Deutsche Forschungsgemeinschaft (DFG) under grant agreement No. 421613952 (ParaStiChaD).
\clearpage
\bibliographystyle{IEEEtran}

\bibliography{mybib}

\end{document}